\documentclass{icrc29}
\usepackage{graphicx, amssymb,amsmath,times}
\begin{document}
\title[Energy loss of muons and taus ...] {Energy loss of muons and taus through inelastic scattering on nuclei}

\author[S.I. Sinegovsky et al.] {A.A. Kochanov, K.S. Lokhtin, S.I. Sinegovsky \\
        Irkutsk State University, Gagarin Blvd 20, Irkutsk 664003, Russia
                }
\presenter{Presenter: A.A. Kochanov (alexey-irk@mail.ru), \
rus-kochanov-A-abs2-he21-poster}

\maketitle

\begin{abstract}

A hybrid model~\cite{KLS2004} was used to describe the energy loss of very high-energy taus and
muons in matter due to inelastic scattering on nuclei. The model involves soft and semihard
photonuclear interactions as well as the deep inelastic scattering including the weak neutral
current processes. For the lepton scattering on nuclei all important nuclear effects, the
shadowing, anti-shadowing, EMC and the nucleon binding, are taken into account. Approximating
formulas for the muon and tau energy loss portion by the inelastic scattering on nuclei in water
are given for wide energy range up to $10^9$ GeV.

\end{abstract}
\section{Introduction}
The muon inelastic scattering on nuclei contributes noticeably to the energy loss of cosmic rays
muons. The influence of this interaction on the shape of ultra-high energy muon spectra at the
great depth of a rock/water is still unknown in detail. The tau-lepton energy loss is of interest
in view of ability of  the atmospheric or extraterrestrial muon neutrinos to transform to the tau
neutrinos which may in turn produce taus in $\nu N$-interactions.

In Ref.~\cite{KLS2004}, the hybrid (two- and three-component) model was proposed to describe
high-energy interactions of charged leptons with nuclei. Calculations of differential cross
sections for lepton-nucleon inelastic scattering at the HERA energies were checked making a
comparison with H1 and ZEUS measurements of electron and positron scattering on protons. With this
model muon and tau energy loss spectra due to lepton-nuclear inelastic scattering were computed as
well as the energy loss rate for leptons passing through standard rock. Now we apply the
two-component version (2C) of the model~\cite{KLS2004} to study the energy loss in the scattering
on nuclei of very-high energy muons and taus passing through water or rock.
The difference in inelastic scattering of the oppositely charged leptons that might originate from
the weak neutral current processes is considered.

\section{Charged lepton inelasic scattering on the nuclei}
The  hybrid two-component (2C)  model for inelastic interactions of high-energy muons and taus with
nuclei involves photonuclear interactions at low and moderate momentum transfer squared as well as
the deep inelastic scattering (DIS) processes at high $Q^2$. For virtuality $0<Q^2<5$ GeV$^2$ the
Regge based parametrization~\cite{CKMT, KMP} for the electromagnetic structure function
$F_2^{\gamma}$ was applied and the lepton-nucleon inelastic scattering cross section at $Q^2\leq 5$
GeV$^2$ was computed with the formula
\begin{equation}
\label{CS_CKMT} \frac{d^2\sigma}{dQ^2dy}=\frac{4\pi\alpha^2}{yQ^4}\left[1-y-\frac{Q^2}{4E^2}
+\frac{y^2}{2(1+R)}\left(1-\frac{2m_\ell^2}{Q^2}\right)
\left(1+\frac{Q^2}{E^2y^2}\right)\right]F_2^{\gamma}(x,Q^2),
\end{equation}
where the ratio $R=\sigma_L/\sigma_T$ is taken into account according to Ref.~\cite{Abe};
$x=Q^2/(2MEy).$   In the DIS range the cross section for the scattering of nonpolarized lepton on
nonpolarized nucleon can be written  in the form~\cite{PDG04}
\begin{equation}
\label{dS_dydQ2}\frac{d^2\sigma}{dQ^2dy}=\frac{4\pi\alpha^2}{yQ^4}
\left[\left(1-y-\frac{Q^2}{4E^2}+\frac{y^2}{2}-\frac{y^2m_\ell^2}{Q^2}\right)
F_2^{NC}\pm\left(\frac{y^2}{2}-y\right)xF_3^{NC}\right],
\end{equation}
where we put  $R=Q^2/(Ey)^2=4M^2x^2/Q^2,$ that is equivalent to the Callan-Gross
relation, $F_2=2xF_1;$ signs ``$\pm$'' stand for $\ell^{\pm}$ ($\ell=\mu,  \tau$).
In Eq. (\ref{dS_dydQ2}) used notations are:
\begin{equation}
\label{F2_NC} F_2^{NC}=F_2^{\gamma}-g_V^\ell \eta_{\gamma Z}F_2^{\gamma
  Z}+({g_V^\ell }^2+{g_A^\ell }^2)\eta_{\gamma Z}^2 F_2^Z, \quad
 F_3^{NC}=-g_A^\ell \eta_{\gamma Z}F_3^{\gamma Z}+
 2g_V^\ell g_A^\ell \eta_{\gamma Z}^2 F_3^Z;
\end{equation}
\begin{equation}\label{eta} \eta_{\gamma Z}=\frac{G_FM_Z^2}{2\sqrt{2}\pi\alpha}
 \frac{Q^2}{M_Z^2+Q^2}, \qquad
 g_V^\ell =-\frac{1}{2}+2\sin^2\theta_W,\qquad g_A^\ell =-\frac{1}{2},\end{equation}
 where $G_F$ is  the Fermi coupling constant, $M_Z$ is the $Z^0$ mass, $\theta_W$  is
the weak-mixing angle.
The structure functions $F_2^Z$, $F_3^Z$  represent the weak neutral current (NC) contribution and
$F_2^{\gamma Z}$,   $F_3^{\gamma Z}$  take account of the electromagnetic and weak current
interference. The nucleon structure functions, $F_2^\gamma$, $F_2^{\gamma Z}$, $F_2^{Z}$,
$F_3^{\gamma Z}$, $F_3^Z$,  are defined in the quark-parton picture:
\begin{equation}
\label{parton_sf} \left[F_2^\gamma,F_2^{\gamma Z},F_2^Z\right]=x\sum_q
\left[e_q^2,2e_qg_V^q,{g_V^q}^2+{g_A^q}^2\right](q+\overline q), \quad
\left[F_3^{\gamma Z},
F_3^Z\right]=\sum_q\left[2e_qg_A^q,2g_V^qg_A^q\right](q-\overline q),
\end{equation}
where $g_V^q=\pm\frac{1}{2}-2e_q\sin^2\theta_W$,  $g_A^q=\pm\frac{1}{2}$. Sign $+\,(-)$ corresponds
to   $u, c, t \, (d, s, b)$-quarks. For the range of $Q^2>5$ GeV$^2$ the electroweak nucleon SFs
are computed with the CTEQ6~\cite {CTEQ6} and the MRST2002~\cite{MRST-02} sets of  the parton
distributions. Linear fits for the nucleon SFs are used in $5< Q^2<6$ GeV$^2$ range. For the
scattering on nuclei the nucleon shadowing, anti-shadowing as well as EMC effect are taken into
account according to Ref.~\cite{smirnov95, smirnov99} (see also~\cite{BM} and \cite{Arneodo94} for
details).
\section{Some results}
\begin{figure}[t]
\begin{center}
\includegraphics[width=7.5cm]{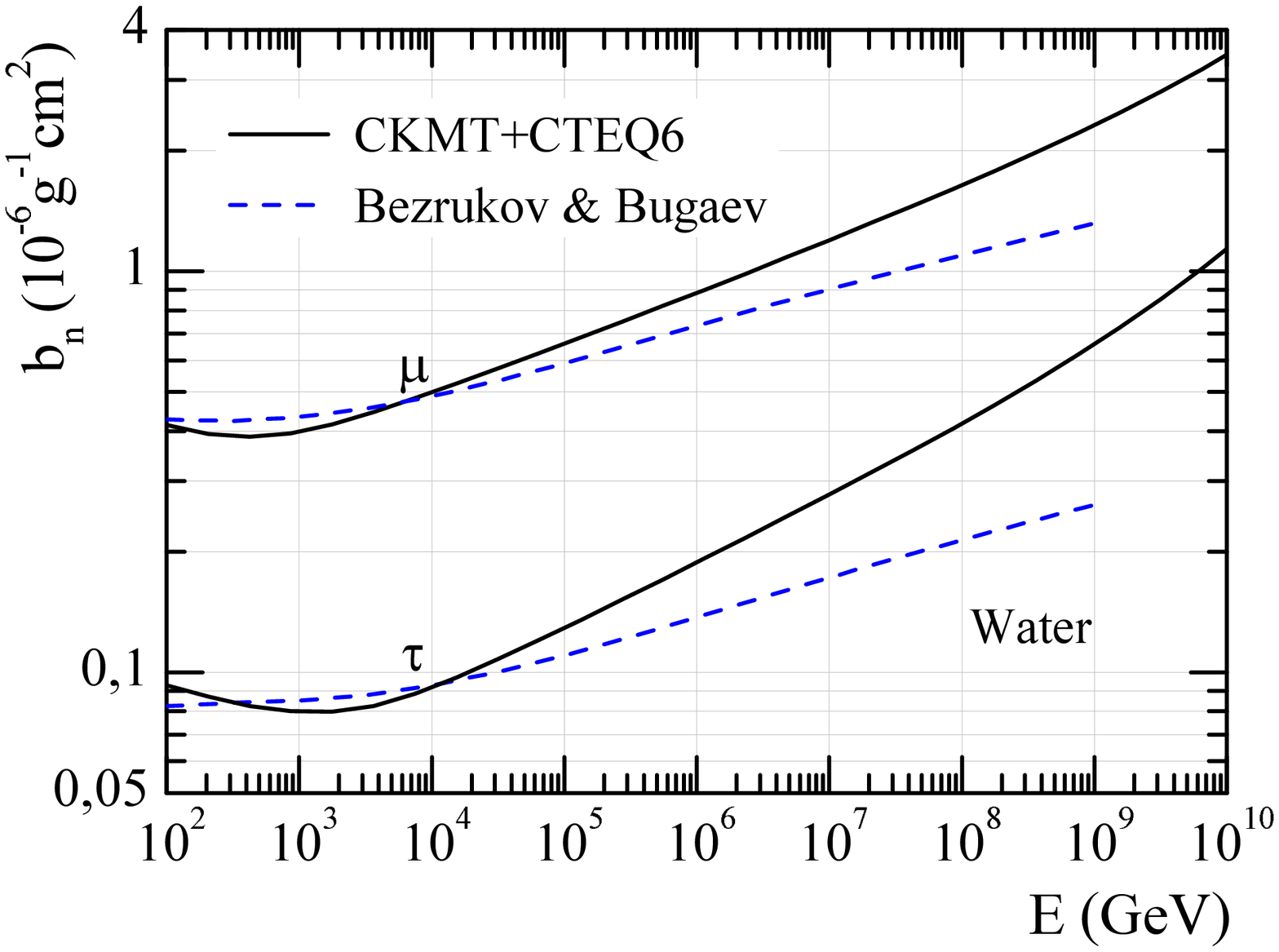} \hskip 0 mm
\includegraphics[width=7.5cm]{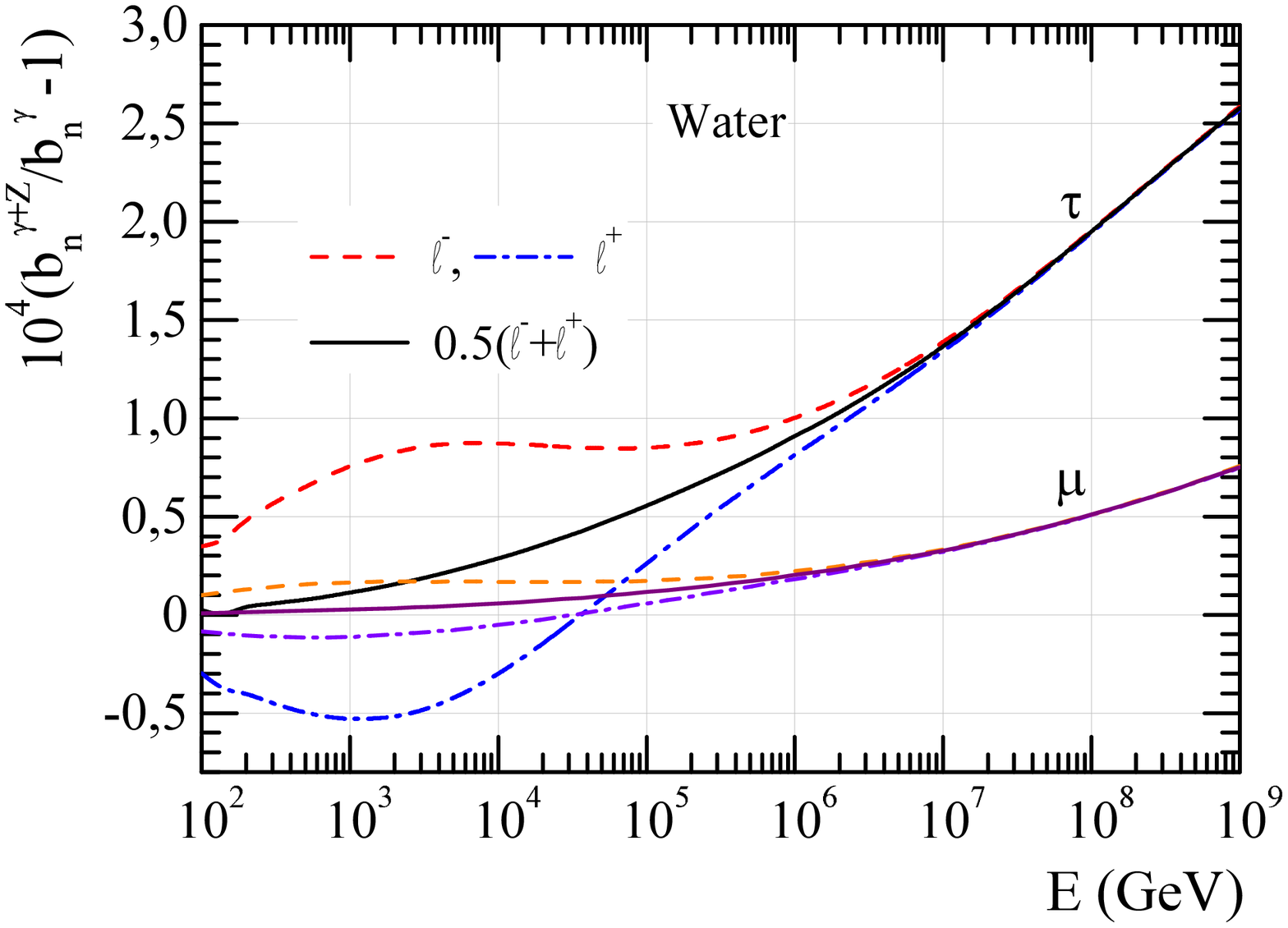}
\end{center}
\vskip -52 mm \caption{Left panel: Energy loss due to the inelasic scattering in water of muons
(upper lines)  and taus (bottom). Right panel: NC contribution to the lepton energy loss due to
$\ell^\pm$-nucleus inelastic scatering in water.} \label{delta-bw}
\end{figure}

The energy loss spectra for lepton passing a substance with nuclear weight $A$ can be derived from
the differential cross-section:
\begin{equation}\label{dsA_dy}
N_0y\frac{d\sigma^{\ell A}}{dy}=N_0\,y\int^{Q^2_{max}}_{Q^2_{min}}dQ^2\,
 \frac{d^2\sigma^{\ell A}}{dQ^2dy},\quad  y=\frac{E-E'}{E}=\frac{\nu}{E},
 \end{equation}
 where $N_0=N_A/A.$
\begin{figure}[b]  \vskip -3mm
\begin{center}
\includegraphics[width=7.5cm]{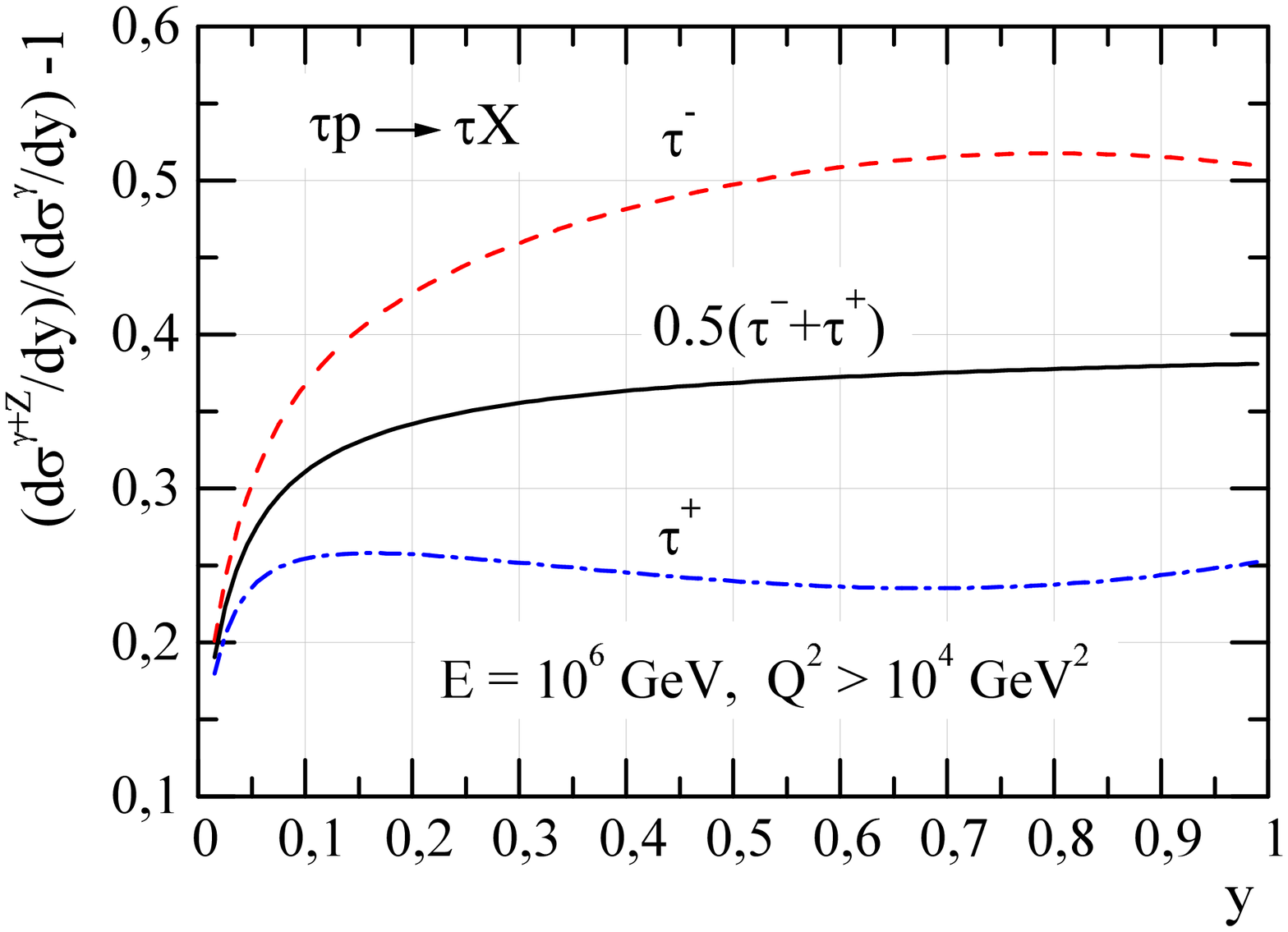}\hskip 2 mm
\includegraphics[width=7.5cm]{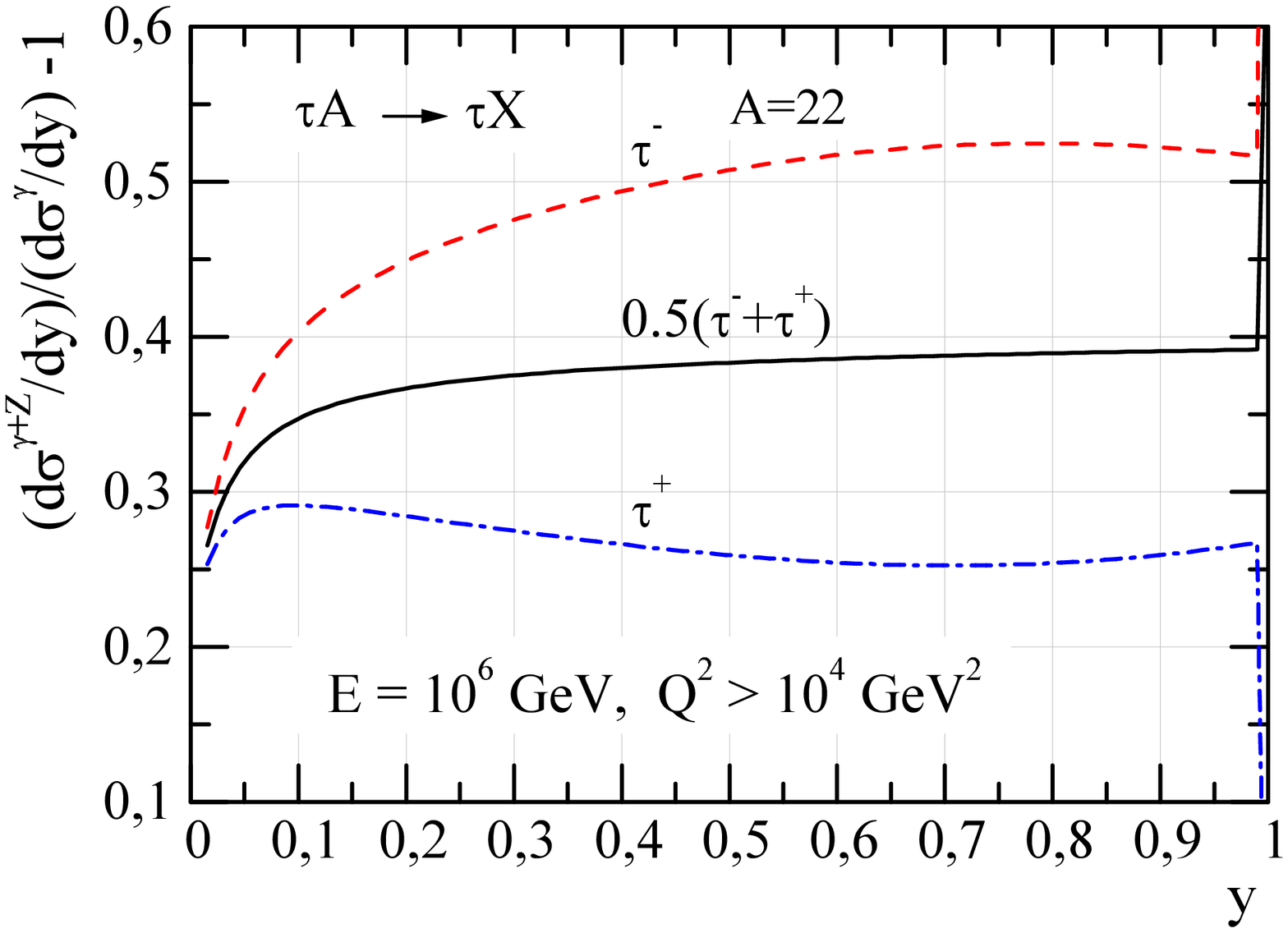}
\end{center}
\vskip -58mm \caption{NC contribution to the cross section of the $\tau^\pm $-proton inelastic
scattering  (left) and that of the $\tau^\pm $-A scattering in standard rock (A=22) (right) at
$E=1$ PeV, \ $Q^2>10^4$ GeV$^2$.} \label{dsdytau_PeV}
\end{figure}
 The energy loss rate due to the lepton-nucleus interactions is defined as
\begin{equation}\label{b_N}
 b^{(\ell)}_n (E) \equiv-\frac 1{E}\frac{dE}{dh}=N_0\,\int^{y_{max}}_{y_{min}}y
 \frac{d\sigma^{\ell A}}{dy}dy. \end{equation}
Figure \ref{dsdytau_PeV} illustrates the charge-dependent NC contribution ($\frac
{d\sigma^{\gamma+Z}}{dy}/\frac {d\sigma^{\gamma}}{dy}-1$) to the lepton  inelastic scattering on
protons (left panel)  or on the nuclei (A=22) (right panel) at $E=1$ PeV, $Q^2>10^4$ GeV$^2.$ The
effect of the  $Z^0$ exchange  is fairly seen just for very  large $Q^2.$ In Fig. \ref{delta-bw},
the left panel, presented are the muon and tau energy loss $b_n(E)$ due to the  inelastic
interactions with nuclei in water (the solid lines). For comparison the  calculations according  to
the vector-meson dominance model~\cite{BB-81} are shown (dashed). The right panel of Fig.
\ref{delta-bw} shows the relative difference of the energy loss,
$(b_n^{\gamma+Z}-b_n^{\gamma})/b_n^{\gamma}$, calculated for the $\ell^+$ and $\ell^-$ particles.
For whole energy range the NC contribution to the $b^{(\ell)}_n (E)$ is too little ($~10^{-4}$) to
be of practical interest.
\begin{table} [h]  
\caption{\label{tab_comp} The muon and tau energy loss in standard rock ($A=22$)}
\begin{center}
\begin{tabular}{c|cccc} \hline \hline
   {  $E,$}& & & ${  b^{(\ell)}_n (E),\  10^{-6}}$ cm$^2\cdot$ g$^{-1}$  \\
  {  GeV}& This work & Ref.~\cite{Dutta} & Ref.~\cite{BM}& Ref.~\cite{BSh} \\\hline
\multicolumn{5}{c}{{Muon}} \\
 $10^5$ &$0.62 $& $0.60$ &$0.68$ & $0.70$ \\
 $10^6$ &$0.82 $& $0.80$ &$0.90$ & $1.08$       \\
 $10^8$ &$1.53 $& $1.50$ &$-$    & $2.25$     \\
 $10^9$ &$2.16$& $2.15$ &$-$    & $3.10$       \\  \hline
\multicolumn{5}{c}{{Tau}}                    \\
 $10^5$  &$0.13 $ & $ 0.12 $ &$-$ & $0.14$   \\
 $10^6$  &$0.19 $ & $ 0.18 $ &$-$ & $0.21$     \\
 $10^8$  &$0.41 $ & $ 0.40 $ &$-$ & $0.50$     \\
 $10^9$  &$0.65 $ & $ 0.60 $ &$-$ & $0.72$    \\  \hline \hline
\end{tabular}
\end{center}
\end{table}

In the second column of the Table~\ref{tab_comp}, the 2C model calculation of the muon and tau
energy loss  in standard rock  and that are presented along with recent
predictions~\cite{BM,Dutta,BSh}. One can see that present calculations of tau-lepton energy loss in
standard rock are similar. As concerns to muons, this work result for $b^{(\mu)}_n$ as well as that
by Ref.~\cite{Dutta} differs apparently at $E>10^6$ GeV from the predictions of Ref.~\cite{BSh}   .

The energy dependence of the inelastic scattering energy loss rate of muons and taus traveling
through water  may be fitted for the range ($10^2-10^9$)  GeV with formula ($\ell=\mu,\tau$):
\begin{equation} \label{fit_b}
 b^{(\ell)}_n (E) =(c_0+c_1\eta+c_2\eta^2+c_3\eta^3+c_4\eta^4)
  \cdot 10^{-6} \ {\rm cm^2/g} \ ,  \  \eta=\lg({E}/1 \,{\rm GeV}),
  \end{equation}
   where coefficients are:\\[-18pt]
\begin{eqnarray}\label{coeff} \nonumber
{\rm \mu:} \ && c_0= 1.06416, \  c_1=-0.64629, \ c_2=0.20394, \ c_3=-0.02465, \
c_4=0.0013;
\\\nonumber
{\rm \tau:} \ && c_0=0.35697, \  c_1=-0.24437, \  c_2=0.07403, \ c_3=-0.00940, \
c_4=0.00051.
\end{eqnarray}



\section{Conclusions}


Recent calculations~\cite{KLS2004}, \cite{BM}, \cite{Dutta}, \cite{BSh} of the energy loss in  the
tau-nuclear interactions are compatible at least for lepton energy up to $10^9$ GeV.  However there
is the discrepancy  between predictions for high-energy behavior of the muon energy loss,
$b^{(\mu)}_n(E)$, in Refs. \cite{Dutta} and \cite{KLS2004}) on the one side, and that of
Ref.~\cite{BSh} on the other side, likely due to diverse ways in
considering of the nuclear effects and high $Q^2$ processes.

The neutral current ($Z^0$ exchange) contribution to energy loss of muons and taus is found to be
negligible  both in water and standard rock on whole energy range (up to $10^{12}$ GeV). Though the
ratio of the cross section for inelastic scattering of $\tau^-$ to that of $\tau^+$ is a sizeable
for $Q^2>10^4$ GeV$^2$ at not too large energies, the effect for the energy loss ($\Delta b_n \sim
10^{-4}\cdot b_n$) seems too small in the cosmic ray physics context.

\section{Acknowledgements}

We acknowledge partial financial support from the Russian Ministry of Education
and Science, grant ur.02.01.063 "Universities of Russia".

\end{document}